\documentclass[aps,prb,twocolumn,nopacs,amssymb]{revtex4}

\usepackage{amsmath}

\usepackage{graphicx}% Include figure files
\usepackage{dcolumn}% Align table columns on decimal point
\usepackage{bm}% bold math
\usepackage{color}

\usepackage{amsmath}
\usepackage{amssymb}

\def\be{\begin{equation}}
\def\en{\end{equation}}
\def\bea{\begin{eqnarray}}
\def\ena{\end{eqnarray}}
\def\p{\partial}
\def\ep{\epsilon}
\def\gs{\gtrsim}
\def\ls{\lesssim}

\newcommand{\av}[1]{\langle{#1}\rangle}

\newcommand{\bi}[1]{\mbox{\boldmath$#1$}}

\newcommand{\ov}[1]{\overline{#1}}

\begin{document}
\title{Global space correlations of polarization, 
charge density, and electric field\\
 in electrolytes  under the 
 fixed-potential condition  
}  
\author{Akira Onuki\footnote{e-mail: onuki@scphys.kyoto-u.ac.jp 
(corresponding author)}}

%\thanks{}
\affiliation{
 Department of Physics, Kyoto University, Kyoto 606-8502, Japan 
}


\date{\today}

\begin{abstract} 
We examine  the thermal 
fluctuations of the polarization 
$\bi p$,  the charge density $\rho$, and the 
electric field $\bi E$   in dilute electrolytes 
 inserted between pararell 
metallic  electrodes, where we  fix   the 
applied potential difference $\Phi_a$ 
 between the two 
electrodes.  If the film  thickness  $H$ 
is shorter than the Debye screening length $\kappa^{-1}$, 
the space correlation of the polarization 
$p_z$ and the electric field $E_z$ 
along   the  surface normal (in  the $z$ direction) 
acuire    global   components  
inversely proportional to the 
film volume $V$, which  vary  slowly along the  $z$ axis  
and  are homogeneous in  the $xy$ plane. 
The  areal charge density 
on each  electrode surface  
  also has  a component homogeneous on the surface, 
which produces  the    global electric   fluctuations. 
On the other hand, if  $H$ much exceeds $\kappa^{-1}$, 
the global correlations of  $p_z$ and $\rho$ become small 
in the bulk region  outside the electric double layers,  
but that of  $E_z$ remains almost unchanged by ions 
in the whole cell 
at fixed  $\Phi_a$. 
The  dielectric constant $\ep_{\rm eff}$ 
depends   on  $H$ and  $\kappa$ 
and is expressed in terms of the 
fluctuation variances of $p_z$ and $\rho$  
and that of the noblocal  surface charge density at fixed $\Phi_a$. 
%correlations  is dominant 
%in the correlation function expression 
%of $\ep_{\rm eff}$. 
\end{abstract}

%\nopacs{}
%\keywords{}

%\maketitle must follow title, authors, abstract, 
%\pacs{ 61.20.Qg, 68.05.Cf, 82.60.Lf, 82.65.Dp }
%and \keywords
\maketitle


\noindent{\bf 1. Introduction}\\
 
Much attention has long 
been paid on the physical and chemical properties 
of electrolytes composed of ions and polar 
fluids\cite{Debye,Landau-e,RobinsonBook}.  
However, many  aspects  of electrolytes  
are still not well understood. 
In this paper, we are interested in 
the fluctuations of the solvent polarization $\bi p$ 
and the ion charge density $\rho$ 
in  electrolytes confined 
between  parallel metallic electrodes.  
For highly polar fluids  in this geometry  (without free ions) 
the dielectric constant sensitively depends on the film 
thickness\cite{Geim,Hau,T2,Takae,P6,Sakuma,Maty1,Laage} 
due to  the Stern  layers on  solid-fluid 
surfaces\cite{Ig,Behrens,Hamann}. 
Furthermore,  statics  and   dynamics  of $\bi p$ 
in such fluids  strongly depend on whether we fix the 
electrode charge $Q_0$ on one electrode surface 
or the applied potential difference $\Phi_a$ between 
the two electrodes\cite{Onuki2025,Sprik}. 
 We note that {\it nonlocal}  electric 
correlations  (independent of space) 
appear at fixed $\Phi_a$ and also 
in the periodic boundary condition.  
It is worth noting that a   number of  
simulations have been performed at  fixed 
$\Phi_a$\cite{T1,Takae,T2,Hau,Cox1,Le,Sprik1, Cox,
P1,P3,P4,P5,P6,P7,Limmer,La,Sato,S1,Wang,Holm}.  
 The periodic boundary condition 
has  been  used in simulations without 
electrodes\cite{Cai,Lebe,Ku,Leeuw,Lada}.  
These  problems  have not been investigated  for 
electrolytes.

For  polar fluids,  
the dielectric constant behaves as   
\be 
\ep_{\rm eff}= 4\pi \av{Q_0} H^2/V\av{\Phi_a}=\ep/
({1+ \ell_{\rm w}/H}). 
\en 
where $\ep$ is the bulk dielectric constant, 
$H$ is the film thickness, $V$ is the film volume,  and 
$\av{Q_0}$ and $\av{\Phi_a}$ 
are the thermal averages of $Q_0$ and $\Phi_a$, respectively. 
Here,   $\ell_{\rm w}$ is a surface electric 
length\cite{Cox1,Takae,Maty1}, which  is 
enlarged for $\ep\gg 1$ and is  
 of order $10$ nm 
for liquid water. The relation (1) holds both 
at fixed $Q_0$ and at fixed $\Phi_a$ 
and   agrees  with an experiment by Geim's group\cite{Geim}.
It is of interest how the above  finite-size 
effect emerges  in electrolytes. 
Here, we should note that 
 the bulk dielectric constant 
$\ep$ itself   decreases due to   
 the hydration of solvent  molecules 
around ions\cite{An1}, 
which is beyond the scope of this paper.

Without ions  we calculated    the polarization 
 correlation function   
 ${ G}_{\alpha\beta}({\bi r}_1, {\bi r}_2)
= \av{\delta p_\alpha({\bi r}_1)\delta p_\beta({\bi r}_2)}/k_BT$ 
($\alpha,\beta=x,y,z$) at fixed $\Phi_a$\cite{Onuki2025}, 
where $\delta{\bi p}={\bi p}- \av{\bi p}$. 
If $ {\bi r}_1$ and ${\bi r}_2$ are in the bulk region 
far from the surfaces, it behaves as 
\be 
{ G}_{\alpha\beta}({\bi r}_1, {\bi r}_2) 
={ G}^{\rm inf}_{\alpha\beta}({\bi r})
-\frac{\chi}{\ep V}\delta_{\alpha z}\delta_{\beta z}
+\frac{\chi\delta_{\alpha z}\delta_{\beta z}}{
(1+ \ell_{\rm w}/H)V}, 
\en  
where ${\bi r}= {\bi r}_1-{\bi r}_2$ and $\chi=(\ep-1)/4\pi$. 
The $z$ axis is perpendicular to the electrode surfaces. 
%This equation   holds for ${\bi r}_1$ and  $ {\bi r}_2$ 
%far from the surfaces. 
For infinite systems $(V\to \infty)$ 
the correlation is  given by\cite{Debye,Landau-e,RobinsonBook}  
\bea 
&&\hspace{-1cm}
{ G}^{\rm inf}_{\alpha\beta}({\bi r}) 
= \chi\delta_{\alpha\beta}\delta({\bi r}) +  
 (4\pi\chi^2/\epsilon) \nabla_\alpha\nabla_\beta \psi_0(r), 
\ena 
where    $\delta({\bi r})$ represents a microscopically 
localized function  with $\int d{\bi r} \delta({\bi r})=1$  
   and $\psi_0(r) 
= (4\pi r)^{-1}$ for $r$ larger than the molecular length. 
Here, ${\sum}_\alpha{ G}^{\rm inf}_{\alpha\alpha}({\bi r}) 
= (2+1/\ep)\chi \delta({\bi r})$ from 
$\nabla^2\psi_0 =-\delta({\bi r})$.     
The second  and third terms in Eq.(2) 
appear for $\alpha=\beta
=z$, being  proportional to $ V^{-1}$ and  
nonlocal. To see their relevance,  
we  integrate Eq.(2) in the $xy$ plane  fixing $z=z_1-z_2$ as   
\bea 
&&\hspace{-10mm}
\int\hspace{-1mm} d{\bi r}_\perp{ G}_{\alpha\beta}({\bi r}_1, {\bi r}_2) 
= \chi \delta(z)[\delta_{\alpha\beta} 
-(1-1/\ep)\delta_{\alpha z}\delta_{\beta z}] \nonumber\\
&&\hspace{15mm} +{\delta_{\alpha z}\delta_{\beta z}}
[-\chi/\ep H+\chi/({H+\ell_{\rm w}})], 
\ena
where ${\bi r}_\perp=(x,y)$ and $L\gg H$. 
See Eqs.(44), (55), and (135)  
in our previous paper\cite{Onuki2025} for the above results. 
Thus. the variance of the total polarization 
$\int d{\bi r}p_z$ along the $z$ axis is  
$Vk_BT\chi/(1+\ell_{\rm w}/H)$    
due to the third  term in Eq.(2). 
At fixed $Q_0$   in the same 
geometry,  no nonlocal term appears and 
the second term in Eq.(4) 
is nonexistent\cite{Ful3,Hansen}.  
%where the fluctuation of  $\int d{\bi r} p_z$ 
%is much suppressed. 
In the periodic boundary condition in 
a cubic cell\cite{Cai,Onuki2025}, 
there appears a nonlocal correlation 
as  ${ G}_{\alpha\beta} =
{ G}^{\rm inf}_{\alpha\beta}+ 
4\pi\chi^2\delta_{\alpha\beta}/3\ep V$, 
leading to   $\int d{\bi r}{ G}_{\alpha\beta} 
= \chi\delta_{\alpha\beta}$. 
On the other hand, in dilute electrolytes\cite{Debye,Landau-e,RobinsonBook},  
  the   charge-charge and charge-polarization 
 correlations  in the limit $V\to \infty$.  
 are written as\cite{OnukiLong}     
\bea 
&& \hspace{-15mm}
\av{\rho({\bi r})\rho({\bi 0})}/ k_BT=
\frac{\ep\kappa^2}{4\pi}\Big[\delta({\bi r})-
\frac{\kappa^2 }{4\pi r}e^{-\kappa r}\Big], \\
&&\hspace{-15mm}
\av{ p_\alpha ({\bi r})\rho ({\bi 0})}/k_BT 
= -\chi  \nabla_\alpha 
\frac{\kappa^2}{4\pi r}e^{-\kappa r},
\ena 
where $\kappa$ is   the Debye wave number 
and $\nabla_\alpha$ is  the $\alpha$ 
 component  of $\nabla$.   The $\psi_0$ 
in Eq.(3) is also  changed to  $e^{-\kappa r}/4\pi r$ for 
large $r$  with ions added. 
Previously,   we further calculated the 
ion-induced long-range correlation   
of the solvent density in infinite systems\cite{OnukiLong}. 
%in addition to the correlations of $\bi p$ and $\rho$   
%in the limit $V\to \infty$\cite{OnukiLong}. 
In this paper,    we shall find {\it global} correlations 
 in confined dilute electrolytes  at fixed $\Phi_a$, which 
vary on the scale of $\kappa^{-1}$ 
along the $z$ axis and are homogeneous in the $xy$ plane.  

The nonlocal   correlations $(\propto V^{-1}$) appear 
generally in equilibrium   under global constraints.  
 In  fluid mixtures,  the space   correlations of  the 
 number densities  have nonlocal parts     
in the canonical and isothermal-isobaric 
 ensembles\cite{Lebo, OnukiP}, 
where  the total particle numbers are fixed, 
These nonlocal parts disappear in the grand canonical enseble. 
The nonlocal or global 
 correlations  can be  amplified  in systems 
with the  electrostatic interactions with electrodes.

The organization of this paper is as follows. 
In Sec.2, we will explain the 
electrical boundary condition of  fixed $Q_0$ and 
that of fixed $\Phi_a$. In Sec.3, we will present the 
free energy at fixed $\Phi_a$, which will be simplified 
in the one-dimensional case in Secs.4 and 5. 
We will then calculate the correlations 
of $\bi p$ and $\rho$ in Secs.6 and 7, those of the electric field 
in Sec.8, and  the charge correlations  in the 
fluid  and on the surface in Sec.9. 
 In Sec.10, the three dimensional correlations will be 
presented.

\vspace{3mm}
\noindent{\bf 2. Electrostatics }\\
%Fig1
\begin{figure}[t]
\includegraphics[scale=1.5]{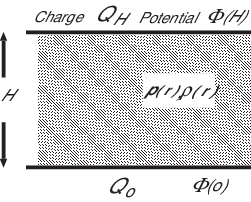}
\caption{\protect  
Electrolyte with polarization  ${\bi p}({\bi r})$ 
and charge density $\rho({\bi r})$ 
between parallel metal plates, where   
$Q_0$ and  $\Phi(0)$ are the electric charge and 
applied potential, 
respectively, at $z=0$, while $Q_H $ and  $\Phi(H)$ 
 are those at $z=H$. 
%The $z$ axis is perpendicular to the plates. 
We set $Q_H=-Q_0$ and $(\Phi_0, \Phi_H)= (\Phi_a,0)$. 
We control $Q_0$  in the fixed-charge condition 
and $\Phi_a$  in the fixed-potential condition.
}
\end{figure}

In the continuum theory, 
we treat  a nearly incompressible 
  polar liquid  with a small amount of ions. 
In this paper, the bulk dielectric constant $\ep$ 
is an arbitrary number larger than 1, while 
we assumed $\ep\gg 1$  in our 
previous paper\cite{Onuki2025}.  
The fluid is confined 
in  a $L\times L\times H$ cell with  volume   $V=L^2H$. 
Here, $L\gg H$ so that  the edge effect is negligible. 
Parallel metallic electrode surfaces  
are  at $z=0$ and $H$ without surface 
chemical reaction. 

 The    polarization  $\bi p$ consists 
of  orientational,  atomic, 
and    electronic parts. 
The  orientational one  is dominant at 
low frequencies in highly polar fluids. 
  The  electric field  ${\bi E}({\bi r})=-\nabla\Phi({\bi r})$ 
is produced by  ${\bi p}({\bi r})$, 
 $\rho({\bi r})$,  and the electrode charges. 
In  the cgs units, the electric induction   
  $ {\bi D}= {\bi E}+4\pi {\bi p}$ satisfies 
   the electrostatic relations\cite{Landau-e},  
\bea 
&&\nabla\cdot{\bi D}= -\nabla^2\Phi+ 4\pi \nabla\cdot{\bi p}
=4\pi\rho,\nonumber~~\\
&&\hspace{-5mm}D_z=4\pi\sigma_0~ (z=0), ~~
D_z=-4\pi\sigma_H~ (z=H), 
\ena 
where $\sigma_0({\bi r}_\perp)$ is the surface 
charge density at $z=0$ and 
$\sigma_H({\bi r}_\perp)$ is that at $z=H$. 
Hereafter, ${\bi r}_\perp=(x,y)$ represents the surface position. 
We assume these relations even for electric thermal 
fluctuations. In Fig.1, we illustrate 
the geometry of our system.

The charge density is written as 
\be 
\rho= Z_1e n_1-Z_2e  n_2=Z_1e \delta n_1-Z_2e  \delta n_2 
\en 
where  $Z_1$ and $Z_2$ are the  valencies and 
$e$ is the elementary charge. 
The  number densities of 
the cations and the anions 
are $n_1=n_{10}+\delta n_1$ 
and   $n_2=n_{20}+\delta n_2$,
 respectively. Here, 
$n_{10}$ and $n_{20}$ are the 
homogeneous densities in equilibrium 
with $  Z_1 n_{10}= Z_2n_{20}$, while 
 $\delta n_1$ and $\delta n_2$ are 
the deviations.  
The Debye  wave number $\kappa$ becomes 
%is given by 
\be 
\kappa=[4\pi n_0e^2/\ep k_BT]^{1/2}, ~~
n_0= Z_1^2n_{10}+Z_2^2n_{20}
\en 
 where  $\kappa= 3.3 \sqrt{m}/$nm 
for monovalent salts with 
 molality $m$ in liquid water.  
We suppose that $\kappa H$ can be of order 1 for 
 small $m$ and $H$. 

We write the total electric charges on 
 the  surfaces as 
\be  
Q_0= \int d{\bi r}_\perp\sigma_0 ({\bi r}_\perp),
~~~Q_H =  \int d{\bi r}_\perp\sigma_H ({\bi r}_\perp).
\en 
We assume no excess charge in the interior so that  
\be 
\int d{\bi r}\rho ({\bi r})=0,~~~Q_0= -Q_H=L^2{\bar\sigma}_0, .
\en 
%The applied field is $4\pi{\bar\sigma}_o
%=4\pi Q_0/L^2$  at fixed $Q_0$ 
%(see Appendix A)\cite{Kubo1}. 
The  potential $\Phi$ 
 can be set equal to  $ \Phi_a$  at $z=0$ and $0$ at $z=H$. 
Experimentally,  either of $Q_0$ or $\Phi_a$ 
 can be fixed by disconnecting or connecting the 
two electrodes. . 

In this paper, we  account for the effect of 
the Stern layers at liquid-metal 
interfaces\cite{Ig,Onuki2025,Hamann,Behrens,Hau,T2,Takae,P6,P4,P3,P1}, 
whose thickness $d_s$  is  a few $\rm \AA$ but 
the electric field inside the layers is very 
 strong. Without absorption of ions  within the layers, 
$D_z$ is continuous across  the layers and the boundary 
conditions in Eq.(7) hold. 
The resultant surface potential drops 
 are $\Phi_{00}+\sigma_0/C_0$ in the range $0<z<d_s$ and  $
-\Phi_{H0}-\sigma_H/C_H$ in the range $H-d_s<z<H$, 
where $\Phi_{00}$ and $\Phi_{H0}$ 
are  the intrinsic charge-free potential drops 
  and $C_0$ and $C_H$ are 
the surface capacitances\cite{Hamann}. 
%We note that $D_z=E_z+4\pi p_z$ is continuous 
%through the layers, so  we can use Eq.(7) 
%even in the layers.  
The  applied potential difference 
$\Phi_a$ is related to  the total  one 
$\Phi_{\rm tot}$  by $\Phi_a= \Phi_{\rm tot}-(\Phi_{00}-\Phi_{H0})$. 
Here,  $\Phi_{00}=\Phi_{H0}$ 
if the two electrode plates are made of the same metal. 
Then,  
\be 
\Phi_a=HE_a = \int dz E_z+ \sigma_0/C_0 - \sigma_H/C_H , 
\en 
where  $E_a=\Phi_a/H$  
and   the integral $\int dz(\cdots)$ 
 is performed outside the Stern layers in the range 
$d_s<z<H-d_s$. 
%For simplicity, we will neglec 
%$\Phi_{00}$ and $\Phi_{H0}$ and 
Hereafter, we will not write 
 $d_s$  explicitly supposing  
 the limit  $d_s \ll H$.  

We further take the lateral average 
%$\int d{\bi r}_\perp (\cdots)/L^2$   
of  Eq.(12), 
\be 
E_a=\Phi_a/H=\int d{\bi r} E_z/V+ {\bar\sigma}_0/CH, 
\en 
where  $\int d{\bi r}(\cdots)$ 
denotes the integration in the cell 
outside the Stern layers. 
We define  ${\bar\sigma}_0$ and $C$ by 
\be 
{\bar\sigma}_0=Q_0/L^2,~~
1/C=1/C_0+1/C_H.  
\en 
The surface electric length $\ell_{\rm w}$ in Eq.(1) 
is given by\cite{Takae}  
\be 
\ell_{\rm w}= \ep/4\pi C ,
\en 
which is amplified by $\ep$. 
See  Eqs.(28) and (41) below.
For liquid water, we have $\ep\cong 80$ 
and $1/4\pi C\sim 1{\rm \AA}$ 
numerically\cite{Takae,Hau,P6} 
 and experimentally\cite{Hamann}, 
so $\ell_{\rm w}\sim 10$ nm.

\vspace{3mm}
\noindent {\bf 3. Free energy at fixed $\Phi_a$ }\\
 
In the dilute limit of ions, 
we expand the translational free energy density of 
the ions with respect to the deviations $\delta n_i=n_i-n_{i0}$. 
Its  bilinear part  is written as 
\bea 
&&\hspace{-10mm} {f_{\rm ion}}={k_BT} 
% [n_1\ln n_1+ n_2\ln n_2)+{\rm const}
[{(\delta n_1)^2}/{2n_{10}}
+ {(\delta n_2)^2}/{2n_{20}}]\nonumber\\
&&\hspace{-4mm} =\frac{1}{2} {\alpha_D }\Big [
\rho^2+ Z_1Z_2e^2 (\delta n_1+\delta n_2)^2\Big],
\ena 
where the coefficient $\alpha_D$ is related to $\kappa$ in Eq.(9) by  
\be 
\alpha_D= 4\pi /\ep \kappa^2. 
\en 
In  the static properties,  
the ion number density $n_1+n_2$ is not affected   by 
the applied field in the linear order, so we set 
 $\delta n_1+\delta n_2=0$.

At fixed $\Phi_a$,  
 the electrolyte 
free energy  is given by\cite{Takae,OnukiB,OnukiLong}  
\be 
{\tilde{\cal F}}= \int\hspace{-1mm} 
d{\bi r}\Big[ \frac{|{\bi E}|^2}{8\pi} 
 + \frac{|{\bi p}|^2}{2\chi }+f_{\rm ion} \Big]
+ \int \hspace{-1mm}
d{\bi r}_\perp f_s -\Phi_a Q_0 
\en 
where ${|{\bi E}|^2}/{8\pi}$ is the electric energy. 
In the second term, $\int d{\bi r}_\perp(\cdots)$ 
denotes the  integration on 
 the surface  $0<x,y<L$. The surface free energy density  
 $f_s$ is assumed to arise from the Stern  layers as  
\be 
f_s= {{\sigma}_0^2}/{2C_0}
+{{\sigma}_H^2}/{2C_H}
\en 
The last term $-\Phi_a Q_0$ in Eq.(18) 
appears in the fixed potential condition. 

We note that the appropriate free energy 
 at fixed $Q_0$, written as $\cal F$,  is obtained 
without the term $-\Phi_aQ_0$. Namely, ${\cal F}={\tilde{\cal F}}
+\Phi_aQ_0$. To understand  this Legendre transformation
\cite{Landau-e,OnukiD,Limmer,Ben}, we 
slightly change ${\bi E}$, $\bi p$, 
$\rho$, $Q_0$, and $\Phi_a$ by 
 $\delta {\bi E}$, $\delta \bi p$, 
$\delta\rho$, $\delta Q_0$, and $\delta\Phi_a$, respectively. 
 Then,   the electric energy is changed by 
\be 
 \int d{\bi r}{\bi E}\cdot\delta{\bi E}/4\pi=
\Phi_a\delta Q_0 +  \int d{\bi r}
[\Phi \delta\rho -{\bi E}\cdot\delta{\bi p}]. \nonumber 
\en 
Here,  the first surface term in the right hand side 
vanishes in $\cal F$ at fixed $Q_0$ 
and in $\tilde{\cal F}$ at fixed $\Phi_a$. 
Thus, we have 
\be 
(\delta {\cal F}/\delta {\bi p})_{Q_0,\rho}
=(\delta {\tilde{\cal F}}/\delta {\bi p})_{\Phi_a,\rho}
= \chi^{-1}{\bi p}-{\bi E}.\nonumber
\en

\vspace{3mm}
\noindent{\bf 4. Laterally averaged 
one-dimensional quantities }\\

We consider 
the one-dimensional profiles 
of $p_z$ and $\rho$ in applied electric field 
along the $z$ axis in the linear 
regime. To this end, 
we introduce the lateral and cell averages,  
\bea 
&&\hspace{-5mm} 
P(z)=\frac{1}{L^2} \int d{\bi r}_\perp p_z({\bi r}), ~~~
\ov{ P} =\frac{1}{H} \int d{z} P(z),\\
&&\hspace{-5mm} 
E(z)= \frac{1}{L^2}
\int d{\bi r}_\perp E_z({\bi r}), ~~~ 
\ov{ E} = \frac{1}{H}\int d{z} E(z), 
%{\bar\sigma}_0= Q_0/L^2,   Q_H=-Q_0, 
\ena 
Using   the charge density $\rho({\bi r})$, 
we also define the one-dimensional charge density and its 
$z$-integral,  
\bea 
&&\rho_{\rm 1D}(z)= \frac{1}{L^2}  \int d{\bi r}_\perp 
\rho({\bi r}_\perp, z),\\
&& \hspace{-10mm}
R(z)=\int_0^z dz' \rho_{\rm 1D}( z'), ~~~
\ov{R} =   \frac{1}{H} \int d{z} R(z),
\ena 
where $R(0)= R(H)=0$ from Eq.(11). Along the $z$ axis,     
\be 
-V\ov{R}= \int d{\bi r} z \rho({\bi r}) 
\en 
is the total charge polarization, while $V\ov{P}$ is the 
total molecular polarization. 
In our system,   the variable  
conjugate to the applied field $E_a$ is 
$\delta{\bar\sigma}_0= ({\ov{P}}-\ov{R})/(1+\delta_s)$ 
(see Eqs.(18), (47), and (50)). 
In the linear response theory  for charged particle systems, 
   Kubo\cite{Kubo} used 
$${\cal H}'= -\int d{\bi r}z\rho({\bi r})E_a$$ as the 
perturbation energy induced by the applied electric field 
$E_a$ (see Eq.(5.7) in his paper).

In terms of ${\bar\sigma}_0= Q_0/L^2$ and $E_a= \Phi_a/H$,  
the electrostatic relations in Eq.(7) yield 
\bea 
&&\hspace{-16mm}
{\bar \sigma}_0= {E(z)}/{4\pi} +P(z)-R(z)\nonumber\\
&& \hspace{-11mm}  
={\ov{ E}}/{4\pi} +\ov{P}-\ov{R} .
\\
&&\hspace{-15mm}\ov{E}= E_a- {\bar \sigma}_0/CH ,
\ena 
where Eq.(25) holds for any $z$. 
Thus, ${\bar\sigma}_0$  is a fluctuating variable 
 expressed in terms of $\ov P$ and $\ov R$ as   
%$\bar{\sigma}_0$ 
%and $E(z)$ can  be  expressed in terms of $P$ 
%and $R$ as 
\bea 
&&\hspace{-16mm}
(1+\delta_s) {\bar\sigma}_0=  \ov{P}-\ov{R}
+E_a/4\pi\nonumber\\
&& \hspace{-2mm}
= \frac{1}{V}\int d{\bi r}\Big[p_z({\bi r})+ z \rho({\bi r}
\Big]+ \frac{1}{4\pi}E_a,
%E(z)= 4\pi{\bar\sigma}_0 -4\pi [P(z)-R(z)], 
\ena 
where $E_a/4\pi$ is the  vacuum term. 
In terms of $C$ and $\ell_{\rm w}$  in Eqs.(14) and (15)
we define 
 $\delta_s$   by 
\be 
\delta_s=1/(4\pi CH)= \ell_{\rm w}/\ep H, 
\en 
which is much smaller than 1.  
Then, $E(z)$ depends on the space average 
$\ov{P}-\ov{R}$ as
\be  
E(z)/4\pi=R(z)-P(z)+(\ov{P}-\ov{R}+E_a/4\pi)/(1+\delta_s).
\en  
Here,  the last  term gives rise to the global 
electric coupling at fixed $\Phi_a$, while 
it is  simply ${\bar{\sigma}}_0 = Q_0/L^2$ at fixed 
$Q_0$ from Eq.(25) leading to no nolocal correlations.   
In this paper, we do not omit    $\delta_s$  
 in $1+\delta_s$ in the course of our  calculations. 
In our final expressions, it  can be safely omitted   
 in $1+\delta_s$ (as in  Eqs.(79) and (91)),    
but   it  should be retained  
in  $1+\ep\delta_s=1+\ell_{\rm w}/H$ for 
$\ep\gg 1$.

The contribution from  $P(z)$ and $R(z)$ to $\tilde{\cal F}$ 
in Eq.(18) is written as $L^2{\tilde{\cal F}}_{\rm 1D}$. 
Since $\rho_{\rm 1D}= dR/dz$, we obtain 
\be 
%\hspace{-10mm}
{\tilde{\cal F}}_{1D} 
\hspace{-1mm} =\hspace{-1mm}
\int \hspace{-1mm}
dz\Big[\frac{E^2}{8\pi} + 
\frac{ P^2}{2\chi }+\frac{\alpha_D}{2}\Big(\frac{d R}{dz}\Big)^2
- E_a {\bar\sigma}_0+ \frac{{\bar\sigma}_0^2}{2CH}
\Big].
\en 
which is of the Ginzburg-Landau form 
with the gradient term $\propto (dR/dz)^2$. 
Note that the polarization gradient 
term $\propto (dP/dz)^2$ is also needed near 
the ferroelectric transition in solids\cite{Binder}, 
where $P$ is the order parameter.

\vspace{3mm}
\noindent{\bf 5. Equilibrium profiles and dielectric constant }\\

We minimize ${\tilde{\cal F}}_{1D}$ in Eq.(30) 
with respect to  $P(z) $ and $R(z)$ 
at fixed $\Phi_a$ to find  their 
 equilibrium averages ${P_e(z)}$ and ${R_e(z)}$. To seek them,  
  we superimpose small deviations   on the variables. 
Then, ${\tilde{\cal F}}_{1D}$ changes by 
\bea 
&&\hspace{-13mm} 
\delta {\tilde{\cal F}}_{1D} = 
\hspace{-1mm}
\int \hspace{-1mm}
dz\Big[\frac{E}{4\pi}\delta E + 
\frac{ P}{\chi }\delta P-{\alpha_D}\frac{d^2R}{dz^2}
\delta R - E\delta{\bar\sigma}_0
\Big]\nonumber\\
&& \hspace{-5mm} = 
\hspace{-1mm}
\int \hspace{-1mm}
dz\Big[\Big(\frac{ 1}{\chi }P-E\Big) \delta P
+\Big(E  -{\alpha_D}\frac{d^2R}{dz^2 }\Big)\delta R\Big].
\ena 
where  we use Eq.(26) and $R(0)=R(H)=0$ 
 in the first line and 
Eq.(25) in the second line. Thus,  we find 
$P= \chi E$ and $ \alpha_D d^2 R/dz^2=E$  at the minimum. 
Setting  $(P,R)=(P_e,R_e)$, we have    
\bea 
&&\hspace{-5mm}
 P_e(z)/\chi=E_e(z)=(4\pi/\ep)  [R_e(z)+ {{\sigma}}_e ] ,\\
&&\hspace{-5mm} d^2 R_e(z)/dz^2=\kappa^{2}[ R_e(z)+ {{\sigma}}_e].
\ena 
Here,   $\sigma_e$ 
is the equilibrium  surface charge density,   
\be 
 {\sigma}_e=\av{{\bar\sigma}_0}
=  (\ep E_a/4\pi-\ov{R_e})/(1+\ep\delta_s).
\en 
where  the product $\ep\delta_s=\ell_{\rm w}/H$ appears.

Using   $R_e(0)= R_e(H)=0$ we solve   Eq.(33) to find  
\bea 
&&\hspace{-10mm} R_e(z)= -{\sigma}_e\psi(z) , \\
&&\hspace{-10mm} P_e(z)= 
 \chi E_e(z)= (4\pi\chi/\ep){\sigma}_e [1-\psi(z)],\\
&&\hspace{-10mm} \rho_{e}(z)=d R_e(z)/dz= 
 {\sigma}_e\kappa \varphi(z) ,
\ena 
where $\rho_e(z)= \av{\rho_{\rm 1D}(z)}$ 
is the equilibrium charge density  
and we define  the functions  
$\psi(z)$ and $\varphi(z)$  by 
\bea 
&&\hspace{-10mm}
\psi(z)=1-\cosh(\kappa z-\kappa H/2)/\cosh (\kappa H/2),\\ 
&&\hspace{-10mm}
\varphi(z)=\sinh(\kappa z-\kappa H/2)/\cosh (\kappa H/2).
\ena 
The average of $\psi(z)$ in the $z$ direction is written as 
\be 
\ov{\psi}=\int_0^H dz \psi(z)/H=
1-2\tanh(\kappa H/2)/\kappa H. 
\en 
We obtain the space averages 
$\ov{R_e}=\int dz R_e(z)/H$ 
and $\ov{P_e}=\int dz P_e(z)/H$ by replacement 
 $\psi(z)\to \ov{\psi}$ in Eqs.(35) and (36). 
For $\kappa H\ll 1$, we have  
$\ov{\psi}\cong (\kappa H)^2/12\ll 1$, 
so  $\psi$, $\ov{\psi}$,  and $\varphi$ 
vanish in the dilute limit $\kappa\to 0$. 
For  $\kappa H\gg 1$, 
we have  $\psi(z)\cong 1- e^{\kappa z-\kappa H}-
e^{-\kappa z}$ and 
 $\ov{\psi}\cong 1- 2/\kappa H$.

%Fig2
\begin{figure}[t]
\includegraphics[scale=0.8]{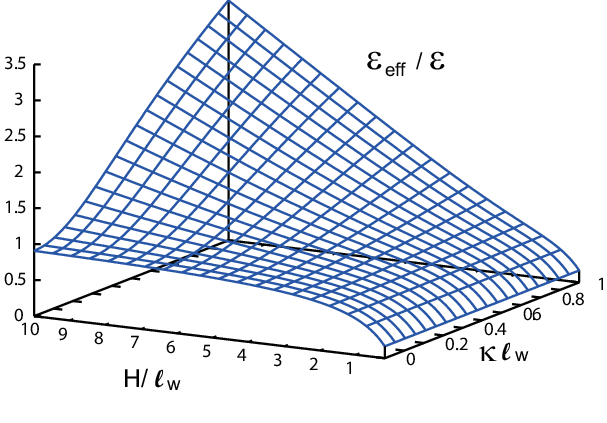}
\caption{\protect  
 Normalized  dielectric constant $\ep_{\rm eff}/\ep$ 
 in Eq.(41) as a function of $H/\ell_{\rm w}$ and 
$\kappa\ell_{\rm w}$ for films with thickness $H$, 
where $\ep$ is the bulk dielectric constant without ions, 
$\kappa$ is the Debye wave number. 
The length $\ell_{\rm w}$ appears in Eqs.(1) and (15) 
 with $\ell_{\rm w}/H= \ep \delta_s$, 
which is about $10~$nm for liquid water.   }
\end{figure}

 From Eqs.(34) and (35) we obtain 
the steady-state    dielectric constant of 
films with ions,  
\bea 
&&\hspace{-6mm}\ep_{\rm eff}= 4\pi {\sigma}_e/E_a=
\ep/(1+ \ep\delta_s- \ov{\psi})\nonumber\\
&&= \ep\kappa H/[\kappa\ell_{\rm w}+ 2\tanh (\kappa H/2)]. 
\ena 
In Fig.2, we display $\ep_{\rm eff}/\ep$ 
as a function of $H/\ell_{\rm w}$ and 
$\kappa\ell_{\rm w}$. For $\kappa H\ll 1$, the above $\ep_{\rm eff}$ 
  tends to the expression (1) for  polar fluids without ions. 
For   $\kappa H\gs 2$,  we have $\ep_{\rm eff}\cong 
 \ep\kappa H/(\kappa\ell_{\rm w}+ 2)\propto H$, 
where the potential 
drops occur mostly  within the 
electric double (ED) layers. 
We note that 
$\ep_{\rm eff}$ in Eq.(41) can be obtained from the 
dynamic theory  by Bazant {\it et al.}\cite{Adjari}  
in the low-frequency limit. 

\vspace{3mm}
\noindent  
{\bf 6. Variable $w$ orthogonal to $R$}\\

We next consider the thermal fluctuations 
of $P$ and $R$ ay fixed $E_a$ and calculate 
their space correlations. 
Their statistical distribution is given by 
\be 
{\cal P}_{\rm dis}(\{P\},\{ R\})= 
{\rm const.}\exp[-L^2{\tilde{\cal F}}_{\rm 1D}/k_BT], 
\en
where ${\tilde{\cal F}}_{\rm 1D}$ is given in Eq.(30) 
and  $\av{\cdots}$ denotes taking the statistical average 
over this distribution. 
In   our theory, this distribution is  Gaussian. 

We introduce a new variable  $w$, 
which is  equal to 
 $P-\chi E$ in the case of $E_a=0$. Some calculations give 
\be 
w= 
\ep P-4\pi\chi R -4\pi\chi(\ov{P}-\ov{R})/(1+\delta_s),
\en 
In particular, the space average of $w$ becomes 
\be 
\ov{w}=\frac{1}{H} \int dz~ w(z) =
\ov{R}+ \frac{1+\ep\delta_s}{1+\delta_s} 
(\ov{P}-\ov{R}). 
\en 
Then,  from Eqs.(25)-(27), 
$P-\chi E$ depends on $E_a$ as 
\be
P-\chi E= \chi \delta{\tilde{\cal  F}_{\rm 1D}}/\delta P
=w-\chi E_a/(1+\delta_s).  
\en  
Thus, the  equilibrium value of $w$ is 
$\chi E_a/(1+\delta_s)$. 
In terms of $w$ and $R$ we  express $P$  as 
\bea 
&&\hspace{-10mm}
 P= \frac{4\pi\chi}{\ep}\Big(R- \frac{\ov{R}}{1+\ep\delta_s}
\Big) + \frac{1}{\ep} (w-\ov{w})+
\frac{1+\delta_s}{1+\ep\delta_s}\ov{w},
\ena
%\be 
%\ov{w}=[({1+\ep\delta_s})\ov{P}}-{4\pi\chi \delta_s}\ov{R} 
%-{\chi E_a}/({1+\delta_s}) 
%\en

We  show that $w$ and $R$ are decoupled in the free energy. 
That is,  ${\tilde{\cal F}}_{1D} $ in Eq.(30) 
can be rewritten  as   
\bea 
&&\hspace{-10mm}
{\tilde{\cal F}}_{1D} 
\hspace{-1mm} =\hspace{-1mm}
\int \hspace{-1mm}
dz ( f_w+ f_R)  -{H}E_a\delta{\bar\sigma}_0,  
\ena 
where  the 
 term quadratic  in $E_a$ is omitted. Here,   
$f_w$ is the free energy density for $w$   
and $f_w$ is that for $R$:     
\bea
&&\hspace{-10mm}
f_w=  \frac{1}{2\chi\ep }(w-\ov{w})^2 
+ \frac{1+\delta_s}{2\chi(1+\ep\delta_s)}{\ov{w}}^2, \\
&&\hspace{-10mm}
f_R= \frac{2\pi}{\ep} R^2  + \frac{\alpha_D}{2}(R')^2
- \frac{2\pi}{\ep(1+\ep\delta_s)} {\ov{ R}}^2.
\ena 
The variable conjugate to the ordering 	 field $E_a$ becomes   
\be 
\delta{\bar\sigma}_0={\bar\sigma}_0-
 \frac{E_a}{4\pi(1+\delta_s)}
=\frac{\ov{P}-\ov{R}}{1+\delta_s}
=\frac{\ov{w}-\ov{R}}{1+\ep\delta_s}.
\en

In Eq.(47) thee is no cross term $\propto wR$ or $\ov{w}\ov{R}$, so  $w$ and $R$ are statistically independent of each other.  
They are  {\it orthogonal} as  
 $\av{R(z)w(z')}=0$  at $E_a=0$, which is obvious from Eq.(45). 
 The correlation of $w$ becomes 
\bea 
&& \hspace{-9mm} \av{w(z)w(z')}\frac{L^2}{k_BT}
= \chi \ep\Big[\delta(z-z')-\frac{1}{H}\Big] 
+ \frac{\chi (1+\ep\delta_s)}{(1+\delta_s)H}\nonumber\\
&&\hspace{5mm}= \chi \ep\delta(z-z')-
{4\pi\chi^2}/[{(1+\delta_s)H}].  
\ena 
where  $\av{(w(z)-\ov{w})\ov{w}}=0$.
In the first line, the first and second 
terms are  equal to $\av{(w(z)-\ov{w})(w(z')-\ov{w})}$ and 
 $\av{{\ov{w}}^2}$, respectively,  multiplied by $L^2/k_BT$, 
%$\delta(z)$ is a microscopically localized 
%function of $z$ with $\int dz\delta (z)=1$. 
In the second line, the second term 
 is  homogeneous and is nonlocal. 
At $E_a=0$,  Eqs.(46) and (51) also give 
\be 
\av{P(z)w(z')}L^2/k_BT = \chi\delta(z-z').
\en  
which holds generally from Eq.(45) (also 
for $Q_0=0$)\cite{OnukiLong}.

From Eqs.(42) and (47) the distribution 
${\cal P}_{\rm dis}$ is 
expanded in  the linear response regime as  
\be 
{\cal P}_{\rm dis}= {\cal P}_{\rm dis}^0[ 1+V E_a 
\delta{\bar\sigma}_0/k_BT +\cdots],
\en
where $ {\cal P}_{\rm dis}^0$ is the distribution 
at $E_a=0$. Therefore, the equilibrium averages $R_e(z)=\av{R(z)}$ 
and $P_e(z)=\av{P(z)}$ in Eqs.(35) and (36) are 
expressed as  
\bea 
&&\hspace{-5mm}
R_e(z)= \av{R(z)(\ov{P}-\ov{R})}{VE_a}/[{k_BT(1+\delta_s)}] ,\\
&&\hspace{-5mm}
 P_e(z)= \av{P(z)(\ov{P}-\ov{R})}{VE_a}/[{k_BT(1+\delta_s)} ].
\ena 
From Eq.(27) the effective dielectric constant is given by 
\bea 
&&\hspace{-10mm}
(1+\delta_s) \ep_{\rm eff}=1 + 
4\pi\av{(\ov{P}-\ov{R})^2}V/[k_BT(1+\delta_s)] 
\nonumber\\
&&\hspace{7mm} =
{1} + 4\pi \av{{\bar\sigma}_0^2}(1+\delta_s)V/k_BT,
\ena 
where $1$ in the right hand sides is the vacuum contribution. 
In Eqs.(54)-(56) the correlations are  those  at $E_a=0$ 
in the linear order. They  will  
be confiirmed  in  Eqs.(74)-(76) below. 

The relations (35)-(37) and (41) 
 also hold at fixed $Q_0$ (see Eq.(A3)). 
This is because  they  are the linear response 
relations among the thermal averages. 
However, the correlations at fixed $Q_0$ and those at fixed $\Phi_a$ 
are very different. For example, in  Appendix A, we will obtain 
\be 
4\pi\av{(\ov{P}-\ov{R})^2}V/k_BT
= 1-1/\ep +\ov{\psi}/\ep~~ (Q_0=0). 
\en 
Compariing  Eq.(56) and Eq.(57),  
we find that  the variance 
$\av{(\ov{P}-\ov{R})^2}$ at $\Phi_a=0$ 
is  larger  than that at $Q_0=0$ roughly  
by a factor of $\ep_{\rm eff}$.  
 For   polar fluids without ions,  
Eqs.(56) and (57) hold for  
$4\pi\av{\ov{P}^2}V/k_BT$ (where $R=\ov{\psi}=0$)\cite{Onuki2025,Sprik}, 
 
 \vspace{3mm} 
\noindent
{\bf 7. Correlation functions  of $P$ and $R$}\\

%Fig3
\begin{figure}[t]
\includegraphics[scale=0.8]{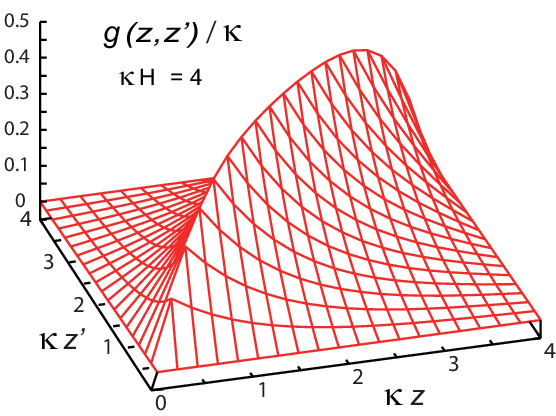}
\caption{\protect  
 Normalized Green function 
$g_\kappa(z,z')/\kappa$ in Eq.(63) 
as a function of $\kappa z$ and $\kappa z'$ 
for $\kappa H=4$. It vanishes if $z$ or $z'$ is 
at $0$ or $H$. It  is given by 
 $\sinh(\kappa z)\sinh(\kappa H-\kappa z)/\sinh(\kappa H)$ 
on  the ridge line  $z'=z$ with 
its maximum being $\tanh (2)/2$.
 }
\end{figure}

%Fig4 
\begin{figure}[t]
\includegraphics[scale=0.8]{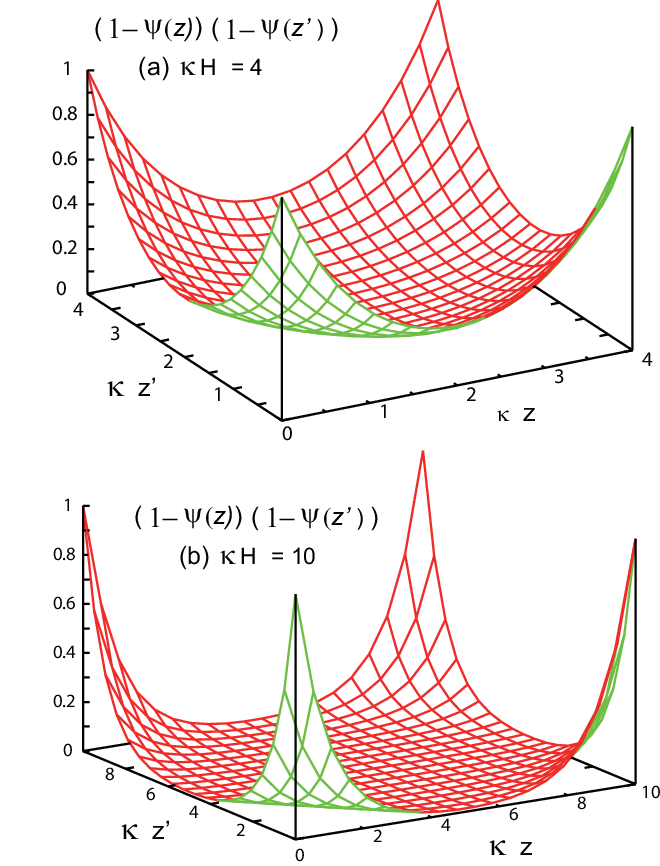}
\caption{\protect  
Two-point factor $(1-\psi(z))(1-\psi(z'))$ 
in  the polarization correlation function $G_(z,z')$
 in Eq.(69), where  $\kappa H$ is 4 in (a) and is  
$10$ in (b). It represents the positive correlation induced by 
the surface charge deviation $\delta{\bar\sigma}_0$. 
 This factor is equal to 1 
for $(x,x')=(0,0),(H,0), (0,H)$, and $(H,H)$. 
Its minimum is equal to $[\cosh(\kappa H/2)]^{-2}$ 
at $(x,x')= (H/2,H/2)$. 
 }
\end{figure}

We now   calculate the space correlations of 
$P$ and $R$ at  $E_a=0$. To simplify  their  
 expressions,  we set 
\bea 
&& \hspace{-5mm}
\av{R(z)R(z')}= ({\ep k_BT}/{4\pi L^2})U(z,z'),\\
&&\hspace{-5mm} \av{P(z)P(z')}=(k_BT/L^2) G(z,z'),\\
&&\hspace{-5mm}  \av{P(z)R(z')}=(k_BT/L^2)  C(z,z'),
\ena  

In Appendix B, we will express $U(z,z')$ in Eq.(58)  as 
\be 
U(z,z')=
g_\kappa(z,z')+ \frac{\psi(z)\psi(z')}{H(1+\ep\delta_s-\ov{\psi})}, 
\en 
where  $\psi(z)$ and $\ov{\psi}$ are 
 given in Eqs.(38) and (40) and $1/(1+\ep\delta_s-\ov{\psi})
=\ep_{\rm eff}/\ep$ from Eq.(41).   
The second term appears due to the global constraint (25) and 
is nonexistent at fixed $Q_0$ (see Appendix A).
The first term $g_\kappa(z,z')$  is   the Green 
function\cite{T1,Doi}, which satisfies   
\be 
(1-\kappa^{-2}\nabla_z^2) g_\kappa(z,z')= \delta(z-z'), 
\en 
with $g_\kappa(0,z')=g_\kappa(H,z')=0$. 
Some calculations give 
\bea 
&&\hspace{-12mm}g_\kappa(z,z')=g_\kappa(z',z)
= \frac{\kappa}{2}e^{-\kappa|z-z'|} 
- \frac{\kappa}{2\sinh(\kappa H)}\nonumber\\
&&\hspace{-1cm}\times[\cosh(\kappa(z+z'-H))
-e^{-\kappa H} \cosh(\kappa(z-z'))].
\ena 
In Fig.3 we display 
 $g_\kappa(z,z')/\kappa$ at  $\kappa H=4$. The 
 maximum of $g_\kappa(z,z')/\kappa$ 
is $ \tanh(\kappa H/2)/2$ at 
$ z= z'=H/2$ for any $\kappa$.  
For $\kappa H\ll 1$, Eq.(63)  gives  
\be 
g_\kappa (z,z')
\cong \kappa^2[ (z+z'-|z-z'|)/2 - z z'/H].
\en 
For $\kappa H\gg 1$ we have  
$g_\kappa(z,z')\cong \kappa e^{-\kappa|z-z'|}/2$ in the bulk region 
outside the ED  layers, 
which is  the lateral  integral 
$\int d{\bi r}_\perp \kappa^2e^{-\kappa r}/4\pi|{\bi r}-{\bi r}'|$. 
In terms of $g_\kappa$ we can express    $\psi(z)$ and $\ov\psi$  in 
Eqs.(38) and (40) as  
\be 
\psi(z)= \int\hspace{-1mm}
 dz'g_\kappa (z,z'), ~~
\ov{\psi}=\hspace{-1mm}\frac{1}{H}\int\hspace{-1mm}
 dz \int\hspace{-1mm} dz'g_\kappa (z,z').
\en 
%We confirm  $U(z,z')= U(z',z)$ and 
%  $U(0,z')= U(H,z')=0$. 
Integrating  Eq.(61) we find  
\bea 
&& {\av{R(z)\ov{R}}}{V}/ k_BT
= (1+\ep\delta_s)\psi(z)\ep_{\rm eff}/4\pi, \nonumber\\
&&{\av{{\ov{R}}^2}}{V}/ k_BT
= (1+\ep\delta_s){\ov\psi}\ep_{\rm eff}/4\pi. 
\ena  
where ${\av{{\ov{R}}^2}}= \int \hspace{-0.5mm}
d{\bi r}\int\hspace{-0.5mm} d{\bi r}'z z' 
\av{\rho({\bi r})\rho({\bi r}'}/V^2$ from Eq.(24). 

From  Eq.(46) we find  $C(z,z')$ in Eq.(60) and 
${\av{{\ov P}{\ov{R}}}}$ as 
\bea 
&& \hspace{-12mm}
C(z,z')=\av{R'(z)R(z')}(4\pi \chi L^2/\ep k_BT)
\nonumber\\
&&=\chi g_\kappa(z,z') -\frac{\chi \ep_{\rm eff}}{\ep H}
{(1-\psi(z))\psi(z')} ,  \\
&&\hspace{-12mm} {\av{{\ov P}{\ov{R}}}}{V}/ k_BT
= \chi{\ov{\psi}}-\chi 
 (1-{\ov{\psi}}){\ov{\psi}} \ep_{\rm eff}/\ep,  
\ena 
where  we  set $R'(z)= 
R(z)-\ov{R}/(1+\ep\delta_s)$ in  Eq.(67).
  
To calculate $G(z,z')$ in Eq.(59) we further use    
\bea 
&&\hspace{-15mm}
\av{R'(z)R'(z')}(4\pi L^2/\ep k_BT)= g_\kappa(z,z') 
\nonumber\\
&&\hspace{-15mm}  +[
{\ep_{\rm eff}(1-\psi(z))(1-\psi(z'))}
-{\ep}/(1+\ep\delta_s)]/\ep H. \nonumber
\ena 
With  the above relation, Eq.(46) then yields   
\bea 
&& \hspace{-15mm}
G(z,z')= \frac{\chi}{\ep}\delta(z-z') 
+\Big(\frac{4\pi\chi^2}{\ep}\Big) g_\kappa(z,z')
\nonumber\\
&&\hspace{-2mm} 
+\Big(\frac{4\pi\chi^2}{\ep}\Big) 
\Big(\frac{\ep_{\rm eff}}{\ep H}\Big) 
(1-\psi(z))(1-\psi(z')).
\ena 
The last terms in  Eqs.(67)-(69) appear at $\Phi_a=0$ 
and disappear   at $Q_0=0$ (see Appendix A).  
In the dilute limit $\kappa\to 0$, Eq.(69) 
gives  the correlations 
for polar fluids,
\bea 
&& \hspace{-10mm}  \lim_{\kappa\to 0}  
 G(z,z') =\frac{\chi}{\ep} \delta(z-z') -\frac{\chi}{\ep H}
+ \frac{\chi(1+\delta_s)}{(1+\ep\delta_s)H}, \\
&& \hspace{-10mm}  
\lim_{\kappa\to 0}
\av{{\ov{P}}^2} \frac{V}{k_BT}\hspace{-1mm}
=\hspace{-1mm}
\int\hspace{-1mm} dz'\lim_{\kappa\to 0}
 G(z,z')=
\chi\frac{1+\delta_s}{1+\ep\delta_s}.
\ena 
Here, we use 
$4\pi\chi^2\ep_{\rm eff}/\ep^2+\chi/\ep=
\chi(1+\delta_s)/(1+\ep\delta_s)$ 
to derive  Eq.(70).  
Thus, Eq.(70) agrees  with  Eqs.(2)-(4)  
if  $1+\delta_s$ in Eq.(70) is replaced by  1.
On the other hand, for $\kappa H\gg 1$,   the last term in Eq.(69) 
is small   in the bulk region 
outside the ED layers, where      
\be 
G(z,z')\cong  \frac{\chi}{\ep}\Big[ \delta(z-z') 
+2\pi\chi\kappa e^{-\kappa |z-z'|}\Big].\en

In Fig.4,   the two-point factor 
$(1-\psi(z))(1-\psi(z')$  in Eq.(69) 
is  displayed, where $\kappa H$ is 
 4 in (a) and 10 in (b). Its  minimum 
is $1/[\cosh(\kappa H/2)]^2$ at $z=z'=H/2$. 
Thus, in Eq.(69),  the ratio of 
 the third term to the  second term  at $z=z'=H/2$ 
is $8e^{-\kappa H}/[(2+\kappa\ell_{\rm w})$ 
  for $\kappa H\gs 2$.
The  polarization correlation extends 
in the whole region of  $z$ and $z'$  
in (a), while $(1-\psi(z))(1-\psi(z')$  
is appreciable only  near the surfaces in (b).  
As a charcteristic feature, the polarization in one ED  
layer and that in the opposite  one are correlated even for 
$\kappa H\gg 1$  (see 
%at $(z,z')=(0,H)$ and $(H,0)$ 
  Fig.4). 

%Fig5 
\begin{figure}[t]
\includegraphics[scale=0.6]{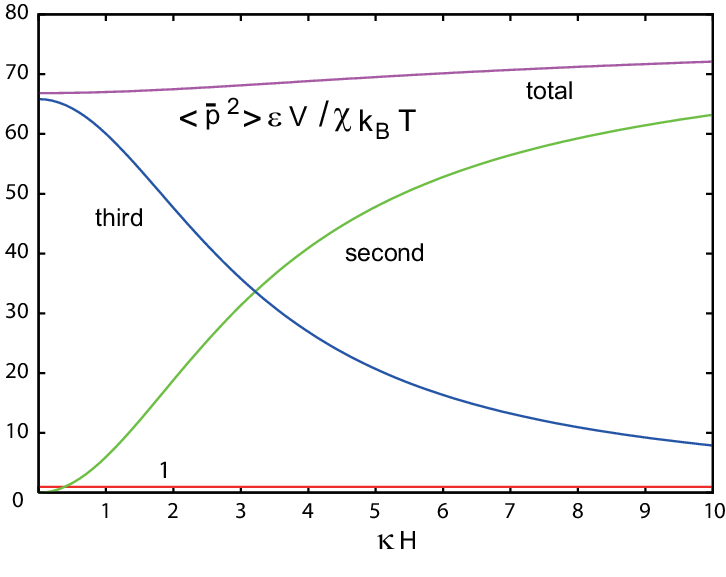}
\caption{\protect  
Three terms and their total sum 
in normalized polarization variance 
$\av{{\ov{P}}^2}\ep V/\chi k_BT$ in the first line of 
Eq.(73) vs  $\kappa H$ 
for liquid water, where  $\ep=80$ and 
$H=5\ell_{\rm w}$. 
At $\kappa=0$ (pure water),  
this quantity is equal to 1 (in red line) 
at $Q_0=0$ and to $\ep/(1+\ell_{\rm w}/H)$ at $\Phi_a=0$. 
For $\kappa >0$, 
it is given by the sum of the second one (in green)  and 1 
at  $Q_0=0$, while it is 
given by the  total sum (in violet)  at $\Phi_a=0$. 
The third term  (in blue)  
appears at $\Phi_a=0$. Since  
  the sum of the second  and  third terms  
is roughly equal to $\ep$, we find  
$\av{{\ov{P}}^2} V/ k_BT\cong \chi$ 
for any $\kappa$ in agreement with  the second line of Eq.(73).  
 }
\end{figure}

We  integrate Eq.(69) over $z$ and $z'$ to find 
\bea 
&& \hspace{-8mm}
\av{{\ov{P}}^2}\ep V/\chi k_BT=
1+ (\ep-1)\ov{\psi}+ (1-1/\ep)\ep_{\rm eff}(1-\ov{\psi})^2 
\nonumber\\
&&\hspace{15mm}= \ep -(\ep-1) (1-\ov{\psi})\ep_{\rm eff}\delta_s,  
 \ena 
where  the   three terms in the first line 
arise from the those in Eq.(69) 
in this order and the second line follows from $
1-(1-\ov{\psi})\ep_{\rm eff}/\ep= \ep_{\rm eff}\delta_s $. 
For liquid water,   Fig.5 shows   the three terms 
  and their sum in the first line of Eq.(73) 
vs  $\kappa H$ with $\ep\delta_s=\ell_{\rm w}/H=0.2$. 
For this $\ep\delta_s$, 
the second line of Eq.(73) 
shows $\av{{\ov{P}}^2} V/k_BT
\cong \chi$   even with  ions.

To confirm the linear response relations 
(54)-(56),  we calculate  the following correlation functions,  
\bea 
&&\hspace{-10mm}
\av{R(z)(\ov{P}-\ov{R})}V/k_BT=
 -(1+\delta_s)\ep_{\rm eff}\psi(z)/4\pi,
\\
&&\hspace{-10mm}
 \av{P(z)(\ov{P}-\ov{R})}V/k_BT=(1+\delta_s) 
\chi \ep_{\rm eff}(1-\psi(z))/\ep,\\
&&\hspace{-10mm}
\av{(\ov{P}-\ov{R})^2}V/k_BT=(1+\delta_s) 
[(1+\delta_s)\ep_{\rm eff}-1]/4\pi, 
\ena 
which lead to  Eq.(54)-(56).

\vspace{3mm} \noindent 
{\bf 8.  Electric field  fluctuations }\\

  We consider  the correlation of the 
one-dimensional electric field 
$E(z)$ in Eq.(21). Since  Eq.(45) gives $E=(P-w)/\chi$ at $E_a=0$,  
it  is calculated as 
\bea 
&& \hspace{-10mm}\av{E(z)E(z')}{\ep L^2}/{4\pi k_BT}
=(\ep-1)  \delta(z-z') + g_\kappa(z,z')\nonumber\\
&&
+ \frac{\ep_{\rm eff}}{\ep H}(1-\psi(z))(1-\psi(z'))
- \frac{\ep}{(1+\delta_s)H}.    
\ena 
%See  Eqs.(A9) and (91) for further calculations. 
Integrating  Eq.(77) over $z'$ and $z$ we find 
\bea  
&&\hspace{-9mm}
{\av{E(z){\ov{E}}}}/{k_BT}
={ \av{{\ov{E}}^2}}/{k_BT}
+{4\pi } \ep_{\rm eff}\delta_s 
[\psi(z)-\ov{\psi}]/V, \nonumber\\
&&\hspace{-9mm}
{ \av{{\ov{E}}^2}}
={ \av{{\bar\sigma}_0^2}}/{C^2H^2}
=16\pi^2\delta_s^2{ \av{{\bar\sigma}_0^2}}, 
\ena  
where we use $1+(\ov{\psi}-1)\ep_{\rm eff}/\ep= 
\ep_{\rm eff}\delta_s$. 
Here,  the fluctuation of the cell average $\ov{E}$ 
  is very small,  
since Eq.(26) gives $\ov{E}= -{\bar{\sigma}}_0/CH
=-4\pi\delta_s{\bar{\sigma}}_0$ at $\Phi_a=0$,

Furthermore, in Eq.(77),  
the fourth term is much larger than the third one 
by a factor of  $\ep\gg 1$ even in the ED layers. Thus, 
for $\ep\gg 1$,  we can approximate Eq.(77) as 
\be 
\frac{\av{E(z)E(z')}}{4\pi k_BT}L^2
\cong \frac{\ep-1}{\ep} 
\delta(z-z') + \frac{g_\kappa(z,z')}{\ep} -\frac{1}{H},  
\en 
where the nonlocal term is simply $-1/H$. 
The $z'$ integrals of the above three terms 
 are $1-1/\ep$, $\psi(z)/\ep$, and $-1$, respectively, 
where their sum  $(\psi-1)/\ep$ is
 much smaller than $1$ for $\ep\gg 1$. At fixed $\Phi_a$, 
the ions hardly suppresses  the nonlocal correlation of 
the electric field in the whole cell  
 even for $\kappa H\gg 1$.
 as it should be the case.

\vspace{3mm} \noindent 
{\bf 9.  Charge fluctuations }\\

The  correlation of the one-dimensional   charge density 
$\rho_{\rm 1D}(z)=dR(z)/dz$ in Eq.(21) is written as 
\bea 
&&\hspace{-10mm}
\av{\rho_{\rm 1D}(z)\rho_{\rm 1D}(z')}{L^2}/{n_0e^2}=
\delta(z-z')-{{h}}_\kappa (z,z') \nonumber\\
&&\hspace{-5mm}
+ ({\ep_{\rm eff}}/{\ep H}) {\varphi (z)\varphi(z')}, 
\ena 
where  $n_0$ is given in Eq.(9),  $\varphi(z)$  in Eq.(38), and 
the $\delta$-function represents  the self correlation. The 
first two terms are those at $Q_0=0$ (see Eq.(A10)), which
 arise from $\kappa^{-2}\nabla_z\nabla_{z'}g_\kappa(z,z')$. 
We calculate $h_\kappa(z,z')$ as 
\bea 
&&\hspace{-12mm}h_\kappa(z,z')
= \frac{\kappa}{2}e^{-\kappa|z-z'|} 
+ \frac{\kappa}{2\sinh(\kappa H)}\nonumber\\
&&\hspace{-1cm}\times[\cosh(\kappa(z+z'-H))
+e^{-\kappa H} \cosh(\kappa(z-z'))].
\ena 
which is   the Green function  satisfying    
\be 
(1-\kappa^{-2}\nabla_z^2) h_\kappa(z,z')= \delta(z-z'), 
\en 
with $\int dz h_\kappa(z,z')=\int dz' h_\kappa(z,z')=1$.

We display  ${{h}}_\kappa(z,z') /\kappa$  
in Fig.6 and the two-point 
factor ${\varphi (z)\varphi(z')}$ in Eq.(80) in Fig.7 for $\kappa H=4$, 
where  the correlation extends in  the cell.  
%As $H\to \infty$, we have 
%$h_\kappa (z,z')\to {\kappa}e^{-\kappa|z-z'|} /2$. 
For $\kappa H\gg 1$, 
the third  term in Eq.(80) is small 
outside the ED  layers so that   
\be 
\av{\rho_{\rm 1D}(z)\rho_{\rm 1D}(z')}
L^2/n_0 e^2
\cong \delta(z-z')-{\kappa}   e^{-\kappa|z-z'|}/2,
\en 
which is the lateral integral of Eq.(5). 
In Eq.(80), however, the third term exceeds the second one close 
to the corners  $(x,x')=(0,H)$ and $(H,0)$, while they 
 are comparable close to  the other corners $(0,0)$ 
and $(H,H)$. 

%Fig6
\begin{figure}[t]
\includegraphics[scale=0.8]{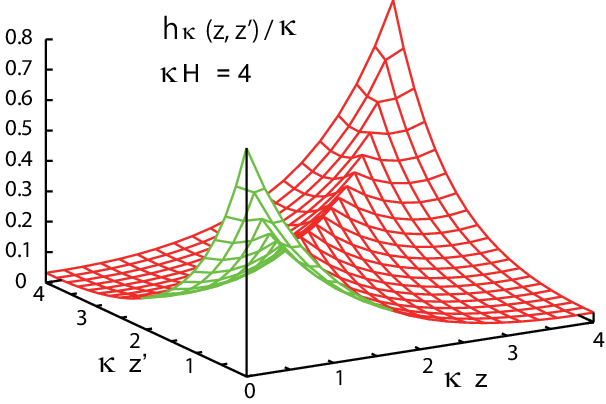}
\caption{\protect  
Normalized Green function $h_\kappa(z,z')$ in Eq.(81) 
at   $\kappa H=4$. It  is equal to 
 $\coth(\kappa H/2)$ at 
$(x,x')=(0,0)$ and $(H,H)$ and to   
 $1/\sinh(\kappa H/2)$ 
at $(x,x')=(0,H)$ and $(H,0)$. 
It has a ridge along the line $z'=z$, on which 
$h_\kappa(z,z)/\kappa =  \cosh(\kappa z)\cosh(\kappa z-\kappa H)/
\sinh(\kappa H)$.  
 }
\end{figure}
%Fig7 
\begin{figure}[t]
\includegraphics[scale=0.8]{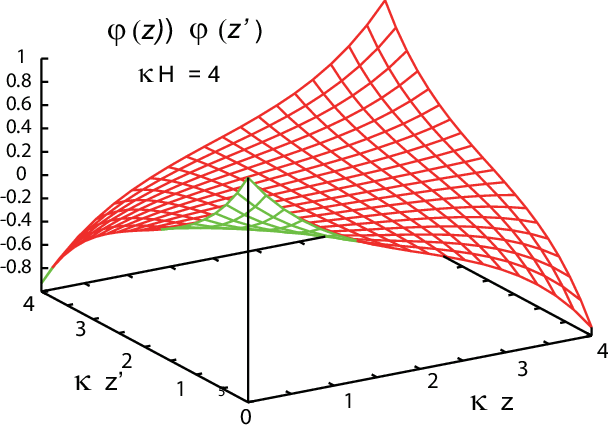}
\caption{\protect  
Two-point factor ${\varphi (z)\varphi(z')}$ in 
the charge correlation function in Eq.(80) at 
  $\kappa H=4$. It represents the correlation induced by 
the surface charge deviation $\delta{\bar\sigma}_0$ 
and is small   outside the ED  layers. 
Its maximum  is $(\tanh(\kappa H))^2>0$ 
at $(x,x')= (0,0)$ and $(H,H)$,  
while its minimum is  $-(\tanh(\kappa H))^2<0$ 
at $(x,x')= (0,H)$ and $(H,0)$.
% The correlation is positive 
%for $(z,z')$ in the same ED layer but is negative 
 }
\end{figure}

The fluctuating surface charge density 
$\sigma_0({\bi r}_\perp)$
at $z=0$ is composed of the thermal average 
$\sigma_e=\ep_{\rm eff}E_a/4\pi$, 
the homogeneous deviation 
$\delta{\bar\sigma}_0= \delta Q_0/L^2 $ in Eq.(50), 
and the inhomogeneous deviation   
$\delta\sigma_{\rm inh}({\bi r}_\perp)$  as 
\be 
\sigma_0({\bi r}_\perp) = \sigma_e + \delta{\bar\sigma}_0 
+\delta\sigma_{\rm inh}({\bi r}_\perp), 
\en 
where ${\bi r}_\perp=(x,y)$. 
At $\Phi_a=0$, Eqs.(50) and (76) give   
\be 
{\av{(\delta{\bar{\sigma}}_0)^2}}
 ={\av{(\delta Q_0)^2}}\frac{1}{L^4 }
=\frac{k_BT}{4\pi V}
\Big[\ep_{\rm eff}-\frac{1}{1+\delta_s}\Big].
\en 
which decreases with increasing $H$ and 
 is small for large $H (\gg \ell_{\rm w})$, However, 
even for large $H$,   ${\delta\bar{\sigma}}_0$ produces 
the global polarization correlation in the cell 
at fixed $\Phi_a$. On the other hand, 
 $\delta\sigma_{\rm inh}({\bi r}_\perp)$ 
satisfies $\int d{\bi r}_\perp
\delta\sigma_{\rm inh}=0$. 
so its in-plane correlation is written as  
\be 
\av{\delta\sigma_{\rm inh}({\bi r}_\perp)
\delta\sigma_{\rm inh}({\bi {\bi r}_\perp'})}
=S_0  \delta ({\bi r}_\perp-{\bi r}_\perp')- S_0/{L^2}.
\en 
where $\delta ({\bi r}_\perp)$ is a localized function 
 with $\int d{\bi r}_\perp \delta ({\bi r}_\perp)=1$ and the 
second  terms  is nonlocal on the surface.
Here, $S_0$ can be  determined from the in-plane 
structure factor,  
\be 
S_{\rm inh}({\bi q}_\perp) 
= \int d{\bi r}_\perp e^{-i{\bi q}_\perp \cdot{\bi r}_\perp}
\av{\delta\sigma_{0}({\bi r}_\perp)
\delta\sigma_{0}({\bi 0})},
\en  
where  $\delta\sigma_0({\bi r}_\perp)= 
\sigma_0({\bi r}_\perp)-\sigma_e$  
and   ${\bi q}_\perp=(q_x,q_y,0)\neq {\bi 0}$. 
Then, $S_{\rm inh}({\bi q}_\perp) \cong S_0$ 
for $\pi/L\ll |{\bi q}_\perp|\ll \xi_\perp^{-1}$, 
where $\xi_\perp$ is the in-plane correlation length.

In our   simulation on pure liquid water\cite{T2}, 
the above in-plane structure factor was calculated, 
where  $\xi_\perp\sim 1{\rm \AA}$ 
 and  $S_0/k_BT\sim 0.03/{\rm \AA}$.
We also found that   the electric field 
perturbations  produced by 
 $\delta\sigma_{\rm inh}$  in the fluid decay rapidly outside 
the Stern layers. 
%the electric field produced by 
%$\delta{\bar\sigma}_0$ extends throughout the cell. 
Previously,  however,  some authors\cite{Limmer,La,P7} 
set    $\av{(\delta {Q}_0)^2}_a/L^2k_BT = S_0$  
neglecting    the in-plane nonlocal fluctuation $\delta{\bar\sigma}_0$.

In the continuum electrostatics,  
the local surface charge densities 
$\sigma_0$ and $\sigma_H$  
are  equal to  $ \pm D_z/4\pi$ at $z=0$ and  $H$ as in Eq.(7), 
which was assumed in     many  
simulations\cite{P1,P3,Hau,T2,Takae}. 
  Siepmann and Sprik\cite{Sprik2} 
 presented an electrode model, where 
atomic particles in the electrodes 
have charges varying continuously 
to realize the metallic boundary condition. 

\vspace{3mm}\noindent 
{\bf 10, Three-dimensional correlation functions}\\ 
%and solvent \\~~~~~~~~~~ density correlation}\\ 

We should consider the  correlation 
functions at $\Phi_a=0$ in three dimensions.  
They consist of those at $Q_0=0$ 
and those produced by ${\bar\sigma}_0$. 
The latter are homogeneous in the $xy$ plane.  
Here,  $\av{\cdots}$ and $\av{\cdots}_{\rm c}$ 
denote the thermal average at $\Phi_a=0$ and 
that  at $Q_0=0$, respectively

From Eqs.(59), (69), and (A8) we find  
\bea 
&&\hspace{-10mm} 
  \av{p_\alpha({\bi r}) p_\beta({\bi r}') }/k_BT =
  \av{p_\alpha({\bi r}) p_\beta({\bi r}') }_{\rm c}/k_BT 
 \nonumber\\
&& \hspace{-8mm} 
+({4\pi\chi^2\ep_{\rm eff} }/{\ep^2}) 
(1-\psi(z))(1-\psi(z'))\delta_{\alpha z}\delta_{\beta z}/V, 
\ena 
which tends to Eq.(2) as $\kappa\to 0$ for  $\delta_s\ll  1$. 
We also present     the other three-dimensional 
correlation functions,   
\bea 
&&\hspace{-11mm} 
\av{\rho({\bi r})\rho({\bi r}')}=
\av{\rho({\bi r})\rho({\bi r}')}_{\rm c}
`+ \frac{n_0e^2\ep_{\rm eff}}{\ep V}{\varphi (z)\varphi(z')} 
\\
&&\hspace{-11mm} 
\av{p_\alpha({\bi r})\rho({\bi r}')}=
\av{p_\alpha({\bi r})\rho({\bi r}')}_{\rm c}
\nonumber\\
&&\hspace{-4mm} 
+ k_BT \kappa  \frac{\chi\ep_{\rm eff}}{\ep V} 
{(1-\psi (z))\varphi(z')}\delta_{\alpha z},
\\
&&\hspace{-11mm} 
\av{E_\alpha({\bi r})E_\beta({\bi r}')}\cong 
\av{E_\alpha({\bi r})E_\beta({\bi r}')}_{\rm c}
-  \frac{4\pi}{V} k_BT\delta_{\alpha z}\delta_{\beta z}.
%&&\hspace{-5mm} \times \frac{4\pi}{\ep V} 
%\Big [ \frac{  \ep_{\rm eff}}{\ep}  
%{(1-\psi (z))(1-\psi(z'))}-\frac{\ep}{1+\delta_s} \Big].
\ena 
In Eq.(91) use is made of  Eq.(79). 
The  correlations $\av{\rho({\bi r})\rho({\bi r}')} $ 
and $\av{p_\alpha({\bi r})\rho({\bi r}')}$ 
tend to those in Eqs.(5) and (6) as $V\to \infty$.  
The lateral integrals of the correlations at $Q_0=0$ 
will be given in Appendix A. 
The  second terms in Eqs.(88)-(91) ($\propto V^{-1}$)
are produced by the surface charge fluctuations, 
 which are significant in the whole cell for $\kappa H\ls 1$. 
For $\kappa H\gg 1$, those in Eqs.(88)-(90)   
 are  small  outside the ED  layers, but that in Eq.(91) 
remains significant. 

%In addition, we consider the solvent density 
%deviation $\delta n_{\rm w}({\bi r})$ 

\vspace{3mm}\noindent 
{\bf 11. Summary and remarks}\\ 

We have studied  static properties of  
 dilute electrolytes between parallel metallic electrodes, 
In particular, fixing  the  potential difference $\Phi_a$ 
between the electrodes at $z=0$ and $H$, 
we have calculated  global space correlations 
of the polarization $p_z$, the charge density $\rho$, and the electric 
field  $E_z$.   
%in the text and those at fixed  eloctrode charge $Q_0$ 
Here, we have three characteristic lengths, the film thickness $H$, 
the surface electric length $\ell_{\rm w}$ due to the Stern layers, 
and the screening length $\kappa^{-1}$, where $\ell_{\rm w}\sim10$nm 
for liquid water. 
For $\kappa H\ls 1$,   the homogeneous part of the surface 
charge fluctuations produces the 
 global   correlations extending in the whole cell, which   vary  
on the scalle of $\kappa^{-1}$ along the $z$ axis and 
 are laterally homogeneous. For $\kappa H\gg 1$,   
the gobal  correlations of $p_z$ and $\rho$  tend to vanish 
in the bulk region outside the electric double (ED) layers, 
but those of $E_z$ still exists in a universal form 
in Eq.(91)  to satisfy the condition of constant  $\Phi_a$. 
Interestingly,  $p_z$ and $\rho$  
in one ED  layer are correlated with 
those in the opposite  ED  layer 
positively for $p_z$ and negatively for $\rho$ 
 even for $\kappa H\gg 1$, as shown in Figs.4 and 7.  
  Furthermore, 
 we have presennted the dielecric constant  in Eq.(41), 
accounting for the sample thickness $H$ and 
the ionic effect, 
the variance of $Q_0$ at $\Phi_a=0 $ in Eq.(83),  
and that of $\Phi_a$ at fixed $Q_0$  in Eq.(A11)

We comment on future problems. 
(1) We should  study  the dynamical global correlations 
of electrolytes in future\cite{Lebe,Onuki2025,Adjari},  
which will be complex even 
in the linear response regime\cite{Kubo}. 
(2) We can also consider the global correlations 
in electrolytes composed of 
mixture solvents and  ions, 
where the fluid can  be close to 
the consolute criticality\cite{OnukiB,Ben}. 
(3) We can also add a dipolar component  
with a large dipole moment into water, 
which tends to accumulate in spatial regions 
with large electric field\cite{OnukiD,An}. 
(4) The global polarization correlations 
emerge most dramatically in systems near the ferroelectric 
transition, which has not been studied.

\vspace{12mm}
\noindent{\bf Appendix A: 
Correlations at fixed $Q_0$ }\\
\setcounter{equation}{0}
\renewcommand{\theequation}{A\arabic{equation}}

As explained in Sec.3, 
the appropriate free energy 
 at fixed $Q_0$ is 
is given by  ${\cal F}={\tilde{\cal F}}
+\Phi_aQ_0$. 
Let  $L^2{{\cal F}_{\rm 1D}}$ 
be the contribution to $\cal F$ 
from $P(z)$ and $R(z)$ 
 in Eqs.(20) and (21) at 
fixed ${\bar\sigma}_0= Q_0/L^2$.  
Here,  the   ordering field  is 
 $E_0\equiv 4\pi {\bar\sigma}_0= 4\pi Q_0/L^2$ 
with  $E_a=\Phi_a/H$ being  a fluctuating variable.  
We assume the electrostatic relation (25). Then, 
we obtain  
\bea 
&&\hspace{-12mm} {{\cal F}_{\rm 1D}}=
 \int \hspace{-1.6mm}
dz\Big[{2\pi} (P-R)^2 +
\frac{1 }{2\chi}P^2 
+ \frac{\alpha_D}{2} \Big(\frac{d R}{dz}\Big)^2\Big]   \nonumber\\
&&\hspace{5mm} 
- H (\ov{P}-\ov{R})E_0 ,
\ena 
where $\ov{P}-\ov{R}$ is the variable conjugate to  $E_0$ 
and the terms quadratic in $E_0^2 $ are not written. 
%Here,   the parameter $\delta_s$ 
%does not appear. 
The statistical distribution of $P$ and $R$ 
is given by const.$\exp(-{{\cal F}_{\rm 1D}}L^2/k_BT)$, 
The averages over this 
 distribution  are written as 
$\av{\cdots}_{\rm c}$.  At fixed $Q_0$, 
there is  no global constraint 
on $P$ and $R$, resulting in  no global correlation. 

We introduce a variable $w$ by   
\be  
w= P-\chi (E- E_0)= \ep P- (\ep-1) R, 
\en   
which is different from $w$ in Eq.(43). 
Since Eqs.(25) and (A2) give 
 $E= E_0 +4\pi(R-w)/\ep$, we have  
\bea 
&&\hspace{-10mm} {{\cal F}_{\rm 1D}}=\hspace{-1mm} \int \hspace{-1.6mm}
dz \Big[\frac{w^2}{2\chi\ep}  + \frac{2\pi}{\ep} R^2 
+ \frac{\alpha_D}{2} \Big(\frac{d R}{dz}\Big)^2\Big] 
\nonumber\\ 
&& -H(\ov{w}-\ov{R})E_0/\ep,  
\ena 
where $w$ and $R$ are decoupled  and  
 $(\ov{w}-\ov{R})/\ep$ is the variable conjugate to $E_0$. 
Minimization of ${{\cal F}_{\rm 1D}}$ gives 
\be 
 w=\chi E_0, ~~~(4\pi/\ep) R-\alpha_Dd^2R/dz^2= 
-E_0/\ep, 
\en     
which then yield  Eq.(35)-(37) and (41). 

At $Q_0=0$, we have   
$\av{w(z)R(z')}_{\rm c}=0$ and   
\bea 
&&\hspace{-8mm} \av{w(z)w(z')}_{\rm c}L^2/k_BT 
= \chi \ep  \delta (z-z'), \\
&&\hspace{-8mm}\av{R(z)R(z')}_{\rm c}L^2/k_BT 
=(\ep/4\pi)  g_\kappa(z,z'), 
\ena 
where there 
arises no nonlocal correlation.
We further use 
 Eq.(A2)  to find  Eq.(52) and 
\bea 
&&\hspace{-10mm}\av{P(z)R(z')}_{\rm c}L^2/k_BT 
=\chi g_\kappa(z,z'), \\
&&\hspace{-10mm}\av{P(z)P(z')}_{\rm c}L^2/k_BT 
=(\chi/\ep)\delta (z-z') \nonumber\\
&& + (4\pi\chi^2/\ep) g_\kappa(z,z'),\\
&&\hspace{-10mm}\av{E(z)E(z')}_{\rm c} L^2/ k_BT 
=(16\pi^2 \chi/\ep)  \delta (z-z')\nonumber\\
&&  +(4\pi/\ep) g_\kappa(z,z'),\\
&&\hspace{-12mm}
\av{\rho_{\rm 1D}(z)\rho_{\rm 1D}(z')}_{\rm c}{L^2}/{n_0e^2}=
\delta(z-z')-{{h}}_\kappa (z,z') ,
\ena 
where $g_\kappa(z,z')$ is given in Eq.(63) and Fig.3, 
 $h_\kappa(z,z')$  in Eq.(81) and Fig.6. and $n_0$ in Eq.(9).

Integrating Eqs.(A5)-(A7)  over $z$ and $z'$ we obtain  
\bea 
&&\hspace{-8mm} \av{{\ov{R}}^2}_{\rm c}V/k_BT= 
\ep\ov{\psi}/4\pi,~~ 
\av{\ov{R}\ov{P}}_{\rm c}V/k_BT= \chi\ov{\psi},\nonumber\\
&& \hspace{-8mm} 
\av{{\ov{P}}^2}_{\rm c}V/k_BT=\chi/\ep + 
(4\pi \chi^2/\ep)\ov{\psi}~~~(Q_0=0),
\ena 
 which lead  to Eq.(57).   
The correspong relations at $\Phi_a=0$ 
are given  in Eqs.(66), (68), and (73). 
At $Q_0=0$, the variance of 
$\Phi_a$ is obtained from Eqs.(25) and (26) as 
\be 
\av{\Phi_a^2}_{\rm c} \frac{1}{H^2}\hspace{-0.5mm}
=\hspace{-0.5mm} 
16\pi^2\av{(\ov{P}-\ov{R})^2}_{\rm c}\hspace{-1mm}
=\hspace{-0.5mm}
\frac{4\pi k_BT}{\ep V} \Big[\ep-1+ \ov{\psi}\Big]. 
\en 
Thus,  $\av{\Phi_a^2}_{\rm c} \cong 
4\pi k_BT H/L^2$ for $\ep\gg 1$  even with addition of  ions.  
The potential time-correlation $\av{\Phi_a(t)\Phi_a(0)}_{\rm c}$ 
will be presented in a forthcoming paper.

\vspace{5mm}
\noindent{\bf Appendix B: 
Correlations of $R$ at fixed $\Phi_a$ }\\
\setcounter{equation}{0}
\renewcommand{\theequation}{B\arabic{equation}}

To calculate   the correlation of $R$, we express 
the space integral of $f_R$ in Eq.(49)  as 
\be
\int dz f_R=\frac{2\pi}{\ep} \int dz\int dz' 
R(z){\cal K}(z,z')R(z'). 
\en 
Here, we introduce  the following two-point 
operator,
\be 
{\cal K}(z,z')=\delta(z-z') -
\nabla_z \frac{\delta(z-z')}{\kappa^2}
\nabla_{z'}
- \frac{1}{H(1+\ep\delta_s)}.
\en 
where $\nabla_z=\p/\p z$ and $\nabla_{z'}=\p/\p z'$.
Then, $U(z,z')$ in Eq.(58) is the {\it inverse} of ${\cal K}(x,x')$ 
satisfying  
\be 
\int dz" {\cal K}(z,z")U(z",z')= \delta(z-z'),
\en  
We thus find  the equation of $U(z,z')$ in the form, 
\be 
\Big(1-\frac{1}{\kappa^{2}}\nabla_z^2\Big)U(z,z')= \delta(z-z')
+ \frac{W(z')}{H(1+\ep\delta_s)}.
\en 
where we set $W(z')= \int dz'' U(z'',z')$. 
The  second term in the right hand side 
is independent of $z$,  leading to 
\be 
U(z,z')= g_\kappa(z,z')+ \psi(z){W(z')}/[{H(1+\ep\delta_s)}],
\en 
where $g_\kappa(z,z')$ is the Green function in Eq.(63) and 
$\psi(z)$ and $\ov{\psi}$ are  given in Eqs.(38), (40), and (65). 
Integration of the above equation 
over $z$ then gives 
\bea 
&&\hspace{-12mm}W(z')= \psi(z')+ \ov{\psi} W(z')/(1+\ep\delta_s)
\nonumber\\
&&= (1+\ep\delta_s)\psi(z')/(1+\ep\delta_s-\ov{\psi}). 
\ena 
We are thus led to  Eq.(61).

\end{document}